\documentstyle[aps,harvard,multicol,epsf,epsfig,eqsecnum,amsmath,amssymb,amsbsy]{revtex}

\renewcommand{\vec}[1]{{\bf #1}}
\citationstyle{dcu}

\def\thefootnote{\fnsymbol{footnote}} 
\begin{document}
\title{The Johnson-Segalman model with a diffusion term in cylindrical
  Couette flow\footnote{To be published in \sl Journal of Rheology}}
\author{ P.~D.  Olmsted, O.
  Radulescu$^1$\footnote{{\tt p.d.olmsted@leeds.ac.uk,
      phyor@irc.leeds.ac.uk, dlu@joule.phy.ncu.edu.tw }} and
  C.-Y.~D.  Lu$^2$ \\
  $^1$Department of
  Physics and Astronomy, and IRC in Polymer Science \\
  and Technology,  University of Leeds, Leeds LS2 9JT, United Kingdom; \\
  $^2$Department of Physics, National Central University, Chung-li, \\
  Taiwan 320, Republic of China} \maketitle
\def\thefootnote{\fnsymbol{footnote}}
\begin{abstract}
  We study the Johnson-Segalman (JS) model as a paradigm for some
  complex fluids which are observed to phase separate, or
  ``shear-band'' in flow. We analyze the behavior of this model in
  cylindrical Couette flow and demonstrate the history dependence
  inherent in the local JS model.  We add a simple gradient term to
  the stress dynamics and demonstrate how this term breaks the
  degeneracy of the local model and prescribes a much smaller
  (discrete, rather than continuous) set of banded steady state
  solutions. We investigate some of the effects of the curvature of
  Couette flow on the observable steady state behavior and kinetics,
  and discuss some of the implications for metastability.
\end{abstract}
\begin{multicols}{2} \narrowtext
\section{Introduction}\label{sec:intro}
\begin{figure}[!tbh] 
\centering{\includegraphics{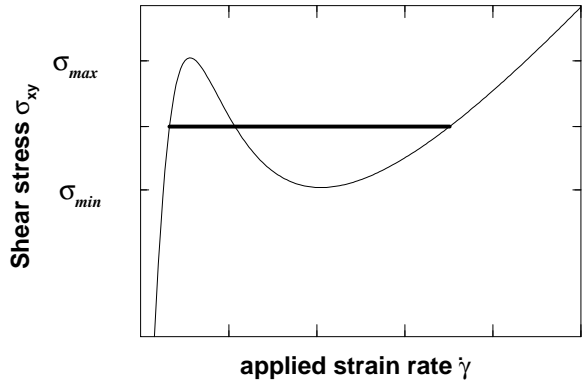}}
\caption{Local constitutive relation
  for the JS model (Eq.~\ref{eq:ssplanar}). The horizontal line shows
  a possible stress for banded flow, with material on either the low
  or high strain rate branches.  The negative slope portion of the
  curve represents mechanically unstable homogeneous flows.}
\label{fig:JS}
\end{figure}

Non-analytic flow behavior is seen in a wide range of complex fluids,
including polymer melts \cite{bagley58,vinogradov72},
wormlike micelles \cite{berret94b,callaghan96,grand97,boltenhagen97b},
and lamellar surfactant phases \cite{roux93,diat93,diat95,sierro97}.
In the ``spurt effect'' in pipe flow
\cite{bagley58,vinogradov72} the flow rate increases
discontinuously above a given applied pressure gradient.  In Couette
flow this behavior manifests itself in non-analyticities in the steady
state flow curves of wall stress as a function of relative cylinder
velocity, with either a discontinuity or a discontinuous first
derivative in the apparent constitutive curve.  In some worm-like
micellar solutions this flow behavior coincides with a range of strain
rates for which macroscopic phase separation is observed
\cite{berret94b,callaghan96,grand97} under controlled strain rate
conditions, with stresses on a well defined, reproducible, and
history-independent plateau.

Although wall slip or internal fracture, rather than bulk instability,
has been suggested to explain the spurt effect \cite{denn90}, bulk
phase separation is undisputed in the wormlike micelle system.  This
phenomenon has often been interpreted in terms of a non-monotonic
constitutive curve analogous to a van der Waals loop, as in
Fig.~\ref{fig:JS}.  Here $\sigma_{xy}$ is the shear component of the
fluid stress tensor, $x$ is the coordinate in the flow direction, and
$y$ is in the gradient direction.  There are many kinds of
non-monotonic behavior \cite{porte97,olmstedlu97}, and we focus here
on systems with multiple strain rates for a given stress.
Constitutive curves of this character are found in the
phenomenological Johnson-Segalman (JS) model
\cite{johnson77,malkus90,malkus91}, the Doi-Edwards microscopic
reptation model for polymers \cite{doiedwards} and the similar model
of \citeasnoun{cates90} for wormlike micelles (both augmented with a
stabilizing high shear-rate branch), and other recent models for
branched polymers \cite{pompom}.  These flow curves have regions of
negative slope, $d\sigma_{xy}/d\dot{\gamma}<0$, which are
hydrodynamically unstable.  For applied strain rates in these unstable
regions, a reasonable scenario consistent with experiments is for the
system to break up into bands of different flow rates, each band lying
on a stable part ($d\sigma_{xy}/d\dot{\gamma} > 0$) of the curve.
This possibility has been studied for various models of non-Newtonian
fluids \cite{mcleish86,mcleish87,olmsted92,spenley96,olmstedlu97}.
However, the stress can apparently lie anywhere within a range of
stresses.  This degeneracy of {\sl selected} stresses has engendered
much discussion, and is at odds with the wormlike micelles, which
shear-band at a well-defined stress that is independent of flow
history \cite{berret94b,callaghan96,grand97,boltenhagen97b}.

Most experiments are performed under conditions which impose
inhomogeneous stress.  For example, in pressure-driven (Poiseuille)
pipe flow the steady state shear stress varies linearly with distance
from the center of the pipe, while in cylindrical Couette flow the
stress varies as the inverse of the square of the radius. There is a
growing body of theoretical work addressing shear banding in
inhomogeneous flow conditions; in particular, we are interested in the
relationship between the degeneracy of models such as the local JS
model, and the possible history-dependence in different flow
geometries, and how to resolve this degeneracy.

\citeasnoun{mcleish87} studied the stability and hysteresis of shear
banding in a modified local Doi-Edwards fluid in pipe flow, and found
that normal stress effects should affect the stability of banded
flows.  \citeasnoun{malkus90} and \citeasnoun{malkus91} studied the
local JS model in planar Poiseuille flow and presented a comprehensive
phase-plane analysis of the steady state behavior, showing how a
hysteresis loop resulted from a particular flow history.
\citeasnoun{greco97} studied shear banding in start-up flows of the JS
model in Couette flow, and demonstrated that the curved geometry of
Couette flow produces a simple well-controlled banded flow, in
contrast to planar flow, in which start-up calculations of similar
local models give an uncontrolled number of bands. More recently the
degeneracy of the JS model was studied in cylindrical Couette
\cite{geovla98} and Poiseuille \cite{fyrilas99} geometries, with the
conclusion in each case that selection of a given banded steady state
depends on the initial perturbation.

In this paper we first study the consequences of degeneracy for the
history dependence of banded solutions to the local JS model in
cylindrical Couette flow by imposing several flow histories. We find,
similar to \citeasnoun{malkus91} and \citeasnoun{fyrilas99} with the
JS model in pipe flow, that the apparent flow curves (in this case,
wall stress as a function of relative cylinder, or ``gap'', velocity)
depend on flow history.  We restrict ourselves to four basic flow
histories and demonstrate how to produce a continuum of possible
solutions. Even in the curved geometry, the solution is unstable with
respect to noise in the initial conditions, as in the planar case
\cite{geovla98}.  This is in distinct contrast to the many
experiments which have found a reproducible and unique selected shear
stress in systems as diverse as worm-like micelles and surfactant
lamellar phases.

To explore unique stress selection we add an additional term to the JS
model, which could be thought of as stress relaxation by diffusion of
differently-strained polymer strands \cite{elkareh89,goveas98}.  Such
``gradient terms'' are not usually included in constitutive equations,
but arise naturally in treatments of hydrodynamics of liquid crystals,
binary fluids, and polymer blends, and are expected to be present
(albeit small) even in one component melts. Such terms will be most
important exactly in the highly inhomogeneous interface between shear
bands, and constitute a \emph{singular perturbation} on the equations
of motion, with a profound effect on stress selection
\cite{lu99,radulescu99a}.  Previous workers have used gradient
terms to study stress selection in models for liquid crystals in
planar shear flow \cite{olmsted92,olmstedlu97}, and in a
one-dimensional scalar model for banded flow in micelle solutions by
\citeasnoun{spenley96}.  Although we make no claims that the JS model
accurately approximates these systems, it is a simple model which
possesses the requisite non-monotonic constitutive flow curve and
whose dynamics, as suggested by Berret, is compatible to some aspects
of kinetics experiments of shear banding in solutions of wormlike
micelles \cite{berret94b,Berr97}.

The outline of this work is as follows. In Section~\ref{sec:model} we
present the equations for the JS model in Couette flow, including the
new diffusion term. In Section~\ref{sec:Dzero} we present the results
from numerical experiments for prescribed strain rates (gap velocity)
and different flow histories in the \emph{absence} of diffusion; and
in Section~\ref{sec:Dnonzero} we examine the effect of diffusion,
which renders the model more realistic with respect to the body of
work on complex fluids which display unique stress selection.  We
finish with a discussion. In a companion paper \cite{radulescu99a} we
analyze the diffusive JS model in Poiseuille and Couette geometries,
using asymptotic matching techniques, to analytically support many of
the findings in this work.
\section{Model}\label{sec:model}
\subsection{Dynamical Equations}
The Johnson-Segalman model has been discussed by many authors
\cite{johnson77,malkus90,malkus91,greco97}. The momentum balance
condition for an incompressible fluid is
\begin{equation}
  \label{eq:NS}
  \rho\left(\partial_t + \vec{v}\!\cdot\!\boldsymbol{\nabla}\right)\vec{v}
  = \boldsymbol{\nabla}\!\cdot\!\vec{T},
\end{equation}
where $\rho$ is the fluid density, $\vec{v}$ the velocity field, and
$\vec{T}$ the stress tensor. The stress tensor is given by
\begin{equation}
  \label{eq:sigma}
  \vec{T} = -p\,\vec{I} + 2\eta\vec{D} + \boldsymbol{\Sigma},
\end{equation}
where the pressure $p$ is determined by incompressibility
($\boldsymbol{\nabla}\!\cdot\!\vec{v}=0$), $\eta$ is the ``solvent''
viscosity, and $\boldsymbol{\Sigma}$ is the ``polymer'' stress. We
separate the velocity gradient tensor
$(\boldsymbol{\nabla}\vec{v})_{\alpha\beta}\equiv
\partial_{\alpha}v_{\beta}$ into symmetric and anti-symmetric parts
$\vec{D}$ and $\boldsymbol{\Omega}$, respectively:
\begin{subequations}
  \label{eq:vgrad}
\begin{eqnarray}
  2\vec{D} &=& \boldsymbol{\nabla}\vec{v} +
    \left(\boldsymbol{\nabla}\vec{v}\right)^T \\
  2\boldsymbol{\Omega} &=& \boldsymbol{\nabla}\vec{v} -
    \left(\boldsymbol{\nabla}\vec{v}\right)^T.
\end{eqnarray}
  \end{subequations}
  The non-Newtonian ``polymer'' stress is taken to have the form
  proposed by \citeasnoun{johnson77}:
\begin{equation}
  \label{eq:JS}
  \overset{\blacklozenge}{\boldsymbol{\Sigma}}  = {\cal
    D}\nabla^2\boldsymbol{\Sigma} + 2\frac{\mu}{\tau}\vec{D} - 
  \frac{1}{\tau}\boldsymbol{\Sigma},
\end{equation}
where $\mu$ is the ``polymer'' viscosity, $\tau$ is a relaxation time,
and ${\cal D }$ is the diffusion coefficient.  The difference from
the usual JS model is the added diffusion term. This has been derived
due to dumbbell diffusion by \citeasnoun{elkareh89}, and is also
expected to have non-local contributions due to the spatial extent of
macromolecules \cite{lu99}.  Gradient contributions to the polymer
stress of blends due to the latter effect have been derived by
\citeasnoun{goveas98}.  The time evolution of $\boldsymbol{\Sigma}$ is
governed by the Gordon-Schowalter time derivative \cite{gordon72},
\begin{eqnarray}
  \label{eq:GS}
  \overset{\blacklozenge}{\boldsymbol{\Sigma}}  = \left(\partial_t +
    \vec{v}\!\cdot\!\boldsymbol{\nabla}\right)\boldsymbol{\Sigma} 
  - \left(\boldsymbol{\Omega\Sigma} - \boldsymbol{\Sigma\,\Omega}\right)
  - a \left(\vec{D}\boldsymbol{\Sigma} + \boldsymbol{\Sigma}\vec{D}\right).
\end{eqnarray}
The ``slip parameter'' $a$ was interpreted by Johnson and Segalman as
a measure of the non-affinity of the polymer deformation; {\em i.e. }
the fractional stretch of the polymeric material with respect to the
stretch of the flow field. For $|a|<1$ the polymer ``slips'' and the
steady-state flow curve in planar shear is capable of the
non-monotonic behavior in Fig.~\ref{fig:JS}.

In this work we consider non-inertial flows (zero Reynolds number
limit), for which the momentum balance Eq.~(\ref{eq:NS}) becomes
$\boldsymbol{\nabla}\!\cdot\!\vec{T}=0.$ Although inertial effects
should not affect stationary solutions, they may influence transients
at the very early stages of start-up experiments, for which inertia
cannot be neglected.  \citeasnoun{spenley96} showed that inertial
terms have practically no influence on the later stages of the
dynamics, so at timescales of order the characteristic relaxation time
$\tau$ the transients are correctly described by the non-inertial
dynamics.  Following \citeasnoun{greco97}, we assume a flow field with
azimuthal symmetry, ${\bf v}= v(r,t)\,\boldsymbol{\hat{\theta}}$
between concentric cylinders of radii $R_1<R_2$, in cylindrical
coordinates $\{r,\theta,z\}$.  Integrating the balance condition
yields
\begin{eqnarray}
\eta \dot{\gamma}(r,t) + \Sigma_{r\theta}(r,t) &=& \frac{\Gamma}{r^2}
\label{eq:torque} \\[5truept]
&\equiv&  \sigma (r,t),
\label{eq:torquebis}
\end{eqnarray}
where we define the local strain rate by
\begin{equation}
  \dot{\gamma}(r,t)\equiv r\,\frac{\partial}{\partial
    r}\left(\frac{v}{r}\right),
\end{equation}
the integration constant $\Gamma$ is the torque per cylinder length,
and $\sigma (r,t)$ is the inhomogeneous shear stress inside the
Couette cell gap.  We consider no-slip boundary conditions on the
fluid velocity ${\bf v}$, a fixed outer cylinder, and an inner
cylinder moving at a prescribed gap velocity $V$. This implies
a global constraint
\begin{eqnarray}
\label{eq:gap}
\frac{V}{R_1} = 
\int_{R_2}^{R_1} \dot{\gamma}(r,t)\frac{dr}{r}
\end{eqnarray}
that may be used to find the torque,
\begin{equation}
  \Gamma \frac{R_1^2 - R_2^2}{R_1^2R_2^2} = 2 \left(\frac{\eta V}{R_1} -
    \int_{R_1}^{R_2} \Sigma_{r\theta}\frac{dr}{r}\right).
\label{eq:global}
\end{equation}

\subsection{Rescaled equations}
In cylindrical coordinates the dynamical equations for
$\boldsymbol{\Sigma}$, Eq.~(\ref{eq:JS}), become
\begin{subequations}
  \label{eq:cyl1}
\begin{eqnarray}
  {\cal L }\,\Sigma_{rr} &=& - (1-a)\dot{\gamma}\Sigma_{r\theta} +
  \frac{2{\cal D }}{r^2}\left(\Sigma_{\theta\theta}-\Sigma_{rr}\right) \\
  {\cal L }\,\Sigma_{\theta\theta} &=&
  \phantom{-}(1+a)\dot{\gamma}\Sigma_{r\theta} -
  \frac{2{\cal D }}{r^2}\left(\Sigma_{\theta\theta}-\Sigma_{rr}\right) \\
  {\cal L }\,\Sigma_{r\theta} &=& \frac{\mu}{\tau}\dot{\gamma} - \frac{4{\cal
      D }}{r^2}\Sigma_{r\theta} - \dot{\gamma}\left[
    \frac{1-a}{2}\Sigma_{\theta\theta} -
    \frac{1+a}{2}\Sigma_{rr}\right]
\end{eqnarray}
\end{subequations}
where
\begin{equation}
  {\cal L } \equiv \partial_t + \tau^{-1} - {\cal D }\nabla^2.
\end{equation}
Next we change variables. We introduce the space variable $x\in(0,1)$
used by \citeasnoun{greco97},
\begin{subequations}
\begin{eqnarray}
  r &=& R_1 e^{qx} \\
  q &=& \ln\frac{R_2}{R_1},
\end{eqnarray}
\end{subequations}
and define
\begin{subequations}
  \begin{eqnarray}
    Z &=& \frac{1-a}{2}\Sigma_{\theta\theta} + \frac{1+a}{2}\Sigma_{rr} \\
    W &=& \frac{1-a}{2}\Sigma_{\theta\theta} -
    \frac{1+a}{2}\Sigma_{rr} \\
    S &=& \Sigma_{r\theta}.
  \end{eqnarray}
\end{subequations}
We now introduce the following dimensionless variables, where $\hat X$
is the dimensionless version of $X$:
\begin{subequations}
  \begin{align}
    \hat V &= V \frac{\tau}{q R_1}\sqrt{1-a^2} &
    \hat W &= W \frac{\tau}{\mu} \\[6truept]
    \hat Z &= Z \frac{\tau}{\mu} &
    \hat{\dot{\gamma}} &= \dot{\gamma} \tau \sqrt{1-a^2} \\[6truept]
    \hat S &= S \frac{\tau}{\mu}\sqrt{1-a^2} &
    {\cal \hat D } &= {\cal D } \frac{\tau}{q^2R_1^2} \\[6truept]
    \hat \Gamma &= \Gamma\frac{\tau}{\mu R_1^2}\sqrt{1-a^2}.&
    \hat \sigma &= \sigma\frac{\tau}{\mu}\sqrt{1-a^2}.
  \end{align}
\end{subequations}
After rescaling,  Eqs.~(\ref{eq:cyl1}) become
\begin{subequations}
  \label{eq:cyl2}
  \begin{eqnarray}
    {\cal L }_x \hat Z &=& \frac{4{\cal \hat
    D }q^2e^{-2qx}}{1-a^2}(\hat W + a \hat Z)a \\
    {\cal L }_x \hat W &=&  \hat{\dot{\gamma}}\hat S 
    -\frac{4{\cal \hat{D} }q^2e^{-2qx}}{1-a^2}(\hat W + a \hat Z) \\
    {\cal L }_x \hat{S} &=& \hat{\dot{\gamma}}\left(1-\hat W\right) 
    -\frac{4{\cal \hat D }q^2e^{-2qx}}{1-a^2}\hat S, 
  \end{eqnarray}
\end{subequations}
with
\begin{equation}
 {\cal L }_x \equiv \partial_t + 1 - {\cal \hat D } e^{-2qx}\partial^2_x.
\end{equation}

The local strain rate and torque (Eqs.~\ref{eq:torque},
\ref{eq:global}) may be written as
\begin{eqnarray}
  \epsilon\hat{\dot{\gamma}} &=& \hat\Gamma e^{-2qx} -  \hat S
  \label{eq:gam}\\
  \hat \Gamma &=& \frac{2q}{1-e^{-2q}} \left(\left\langle \hat S
    \right\rangle - \epsilon \hat V\right)  , \label{eq:Gam}
\end{eqnarray}
where
\begin{align}
  \langle \hat S\rangle &= \int_0^1\!\!dx\,\hat S\\
  \hat{V} &= \frac{1}{q}\int_{R_1}^{R_2}\!\!\hat{\dot{\gamma}}\,\frac{dr}{r} 
  = \int_{0}^{1}\!\!dx\,\hat{\dot{\gamma}}, \label{eq:Vhat}
\end{align}
and
\begin{equation}
  \epsilon = \frac{\eta}{\mu}
\end{equation}
is the viscosity ratio. 

We consider the dynamics of flows for prescribed histories of inner
cylinder velocities $\hat V$, by solving
Eqs.~(\ref{eq:cyl2},\ref{eq:gam},\ref{eq:Gam}).  The only parameters
are ${\cal \hat D }$, which sets the length scale of any interfaces;
the slip parameter $a$; the viscosity ratio $\epsilon$; and the
curvature of the Couette cylinders, determined by $q$. We will often
use the relative gap size,
\begin{eqnarray}
  p &\equiv& e^q - 1 \\
  &=&  \frac{R_2-R_1}{R_1}
\end{eqnarray}
instead of $q$.
\subsection{Boundary Conditions}
While the fluid velocity it taken to obey no-slip boundary conditions,
for $\boldsymbol{\Sigma}$ we choose the following boundary condition:
\begin{align}
  \boldsymbol{\nabla}\Sigma_{\alpha\beta}&=0,\\
  \intertext{or} \partial_x \hat W & = \partial_x \hat Z = \partial_x
  \hat S = 0.
\end{align}
For the interpretation of the stress diffusion term as arising from
the diffusion of polymeric dumbbells \cite{elkareh89}, this
corresponds to a zero flux boundary condition.  In a more realistic
model gradient terms are expected to arise from inter- and
intra-molecular interactions as well as center-of-mass diffusion, and
the boundary condition may be a fixed non-zero gradient, or perhaps a
fixed value for the stress.  Clearly the boundary conditions depend on
the detailed physics. In this work the nature of the boundary
condition is expected to play a role in the limit when the interface
touches the wall, so predictions in this limit may be non-universal.
\section{Couette Flow with no diffusion: ${\cal\hat D }=0$}\label{sec:Dzero}
\subsection{Steady State Solutions}\label{sec:ssD}
The steady states and stability of the local JS model ($\hat{\cal
  D}=0$) in planar shear flow have been studied by many authors.
Although some numerical calculations have shown well-defined shear
banding \cite{espanol96,yuan99}, it is generally believed that the
local model supports shear bands at any stress in the non-monotonic
region \cite{renardy,spenley96,geovla98,lu99}.  In the presence of the
diffusion term stress is selected at a well defined value, independent
of history. We will not discuss the case of ${\cal\hat D }\neq 0$ in
planar flow in detail here, but refer the reader to \citeasnoun{lu99}.

\citeasnoun{greco97} performed start-up calculations for ${\cal\hat D
  }=0$ in a cylindrical Couette geometry and found, unlike the 1+1D
planar simulations \cite{spenley96,espanol96}, an apparently
reproducible steady state with two shear bands and a single zero width
interface between bands. Before studying this model in detail, we
present a graphical construction of the steady state banded solutions.

\begin{figure}[!tbh]
\centering{\includegraphics{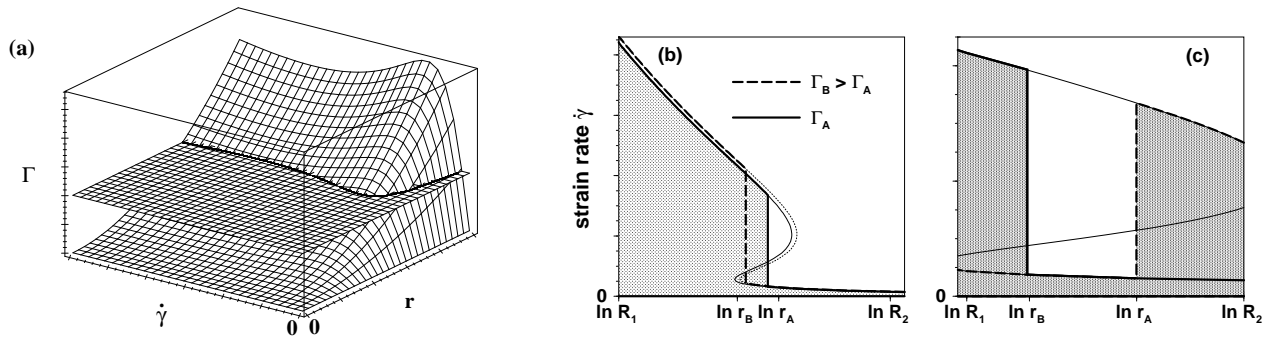}}
\caption{(a) The curved surface shows the 
  torque $\hat\Gamma(r,\dot{\gamma})$ as a function of strain-rate
  $\dot{\gamma}$ and radius $r$ for the Johnson-Segalman model in
  Couette flow, given by Eq.~(\ref{eq:local}).  The plane is at a
  specified torque.  (b,c) Banding profiles at different torques for
  the JS model in Couette flow, with no diffusion term (${\cal\hat
    D }=0$), obtained from the intersection of the plane and curved
  surfaces in (a) according to Eq.~(\ref{eq:local}).  Thin lines are
  strain-rate ${\dot{\gamma}}$ as a function of radius $r$; thick
  lines are banding solutions. (b) Possible profiles for torques
  $\Gamma_A$ and $\Gamma_B>\Gamma_A$ at the same gap velocities $V$,
  given by the area of the shaded region.  (c) Two possible bands for
  a given torque, again with the same gap velocity.}
\label{fig:torque}
\end{figure}

  The steady state solutions to Eqs.~(\ref{eq:cyl2}) for ${\cal\hat
    D }=0$ have the local form as the flat case, given by
  \cite{greco97}
\begin{subequations}
\label{eq:ssplanar}
\begin{eqnarray}
  \hat Z&=&0 \\
  \hat W &=& \hat{\dot{\gamma}} \hat S \\
  \hat S &=& \frac{\hat{\dot{\gamma}}}{1 + \hat{\dot{\gamma}}^2}.
\end{eqnarray}
\end{subequations}
Using the local force balance,
Eq.~(\ref{eq:torque}), and Eq.~\ref{eq:ssplanar}~c) we obtain:
\begin{eqnarray}
\hat\Gamma =  \hat{r}^2
{\hat{\dot{\gamma}}}\left(\epsilon+\frac{1}{1 
+{\hat{\dot{\gamma}}}^2}\right),
\label{eq:local}
\end{eqnarray}
where 
\begin{equation}
  \hat{r} = \frac{r}{R_1}.
\end{equation}
Eq.~(\ref{eq:local}) defines the locus of steady states parametrized
by torque, radial position, and local strain rate.  In the
three-dimensional space spanned by $\{\hat\Gamma, \hat{\dot{\gamma}},
r\}$ this relation defines a surface, shown in Fig.~\ref{fig:torque}a.
Imposing the uniform torque condition (the intersection of the surface
$\hat\Gamma(r,\hat{\dot{\gamma}})$ with a plane) yields a local
relation between strain-rate and radius $\hat{\dot{\gamma}}(r,t)$, shown
in Fig.~\ref{fig:torque}b,c.
\end{multicols} \widetext
\begin{multicols}{2} \narrowtext

A steady state banded flow profile traces a trajectory
$\hat{\dot{\gamma}}(r)$ in Fig.~\ref{fig:torque}b, with a vertical
jump.  In the two solutions shown in Fig.~\ref{fig:torque}b, the high
strain rate phase lies nearest the inner cylinder.  Because of the
inhomogeneity of Couette flow the high and low strain rate phases are
non-uniform.
  
Given a solution, \emph{e.g.} the solid line in Fig.~\ref{fig:torque}b
for torque $\hat\Gamma_A$, the shaded area corresponds to $\hat V$,
according to Eq.~(\ref{eq:Vhat}).  By placing the interface at
different positions, we can generate a continuum of solutions with
different gap velocities for a given torque.  Conversely, it is easy
to find a continuum of torques for a given gap velocity, with each
torque corresponding to a unique interface position.  Upon increasing
the torque to $\hat\Gamma_B>\hat\Gamma_A$ while keeping the interface
fixed (specified by the vertical solid line), the area under the new
$\hat{\dot{\gamma}}(r)$ curve, and hence the gap velocity $\hat{V}$,
increases. To recover the $\hat{V}$ we must move the interface closer
to the inner cylinder ($R_1$), obtaining the solution shown in the
thick dashed line.  Also, multiple interface solutions are possible:
flow profiles which traverse between high and low strain rate branches
an odd number of times are possible steady states.  Hence, for a band
sequence the steady state is completely defined by any two of the
three quantities $\Gamma$, $ V/R_{1}$, and interface position
$r_{\ast}$.  If both the low and high strain branches span the gap
(Fig.~\ref{fig:torque}c), which happens for a small curvature
($R_1/R_2$ close to $1$), both the conventional band sequence, and an
inverted band sequence, with the high strain rate material at the
outer cylinder, are possible.  This possibility is stabilized for
flatter cylinders, and in the planar limit the conventional and
inverted bands are symmetry-related partners.

Hence, a prescribed gap velocity does not select a unique stress
(torque) in cylindrical Couette geometry.  At fixed $ V$, the family
of single interface steady flow solutions is parametrized by $\Gamma$
or by $r_*\in\left[R_1,R_2\right]$.  In principle, a universal banded
steady flow might still exist if all the initial conditions or flow
histories evolved to the same steady state configuration (selected
values of $r_*$ and $\Gamma$).  To explore this we next simulate flows
prepared by different flow histories.

\subsection{Dynamics}
\subsubsection{History dependence}

We have evolved Eqs.~(\ref{eq:cyl2}) through various sequences of gap
velocities $\{\hat V(t_i)\}$, where $t_i$ is the time when the gap velocity
is changed, and considered four types of flow histories:
\begin{enumerate}
\item {\bf Jump-up:} jump each time from a gap velocity $\hat
  V_{down}$, small and on the low strain rate branch, to a gap
  velocity $\hat V_{i}$; $\{\hat V(0)=\hat V_{down},\hat V(t_1)=\hat
  V_{1},\hat V(t_2)= \hat V_{down},\ldots,\hat V(t_{2n})=\hat
  V_{down},\hat V(t_{2n+1})=\hat V_{n}\}$.  In particular $\hat
  V_{down}=0$ means starting from rest.
\item {\bf Jump-down:} jump each time from a velocity $\hat V_{up}$,
  large and always on the high strain rate branch, to a velocity $\hat
  V_{i}$; $\{\hat V(0)=\hat V_{up},\hat V(t_1)=\hat V_{1},\hat V(t_2)=
  \hat V_{up},\ldots,\hat V(t_{2n})=\hat V_{up},\hat V(t_{2n+1})=\hat
  V_{n}\}$.
\item {\bf Ramp-up:} increasing sequence of
  velocity values; $\{\hat V(0)=0 < \hat V(t_1)=\hat V_{1} < \ldots
  < \hat V(t_{n})=\hat V_{n}\}$.
\item {\bf Ramp-down:} decreasing sequence of
  velocity values; $\{\hat V(0)=\hat V_{n} > \hat V(t_1)=\hat V_{n-1}
  > \ldots > \hat V(t_{n-1})=\hat V_{1}\}$.
\end{enumerate}

\begin{figure}[!tbh]
\centering{\includegraphics{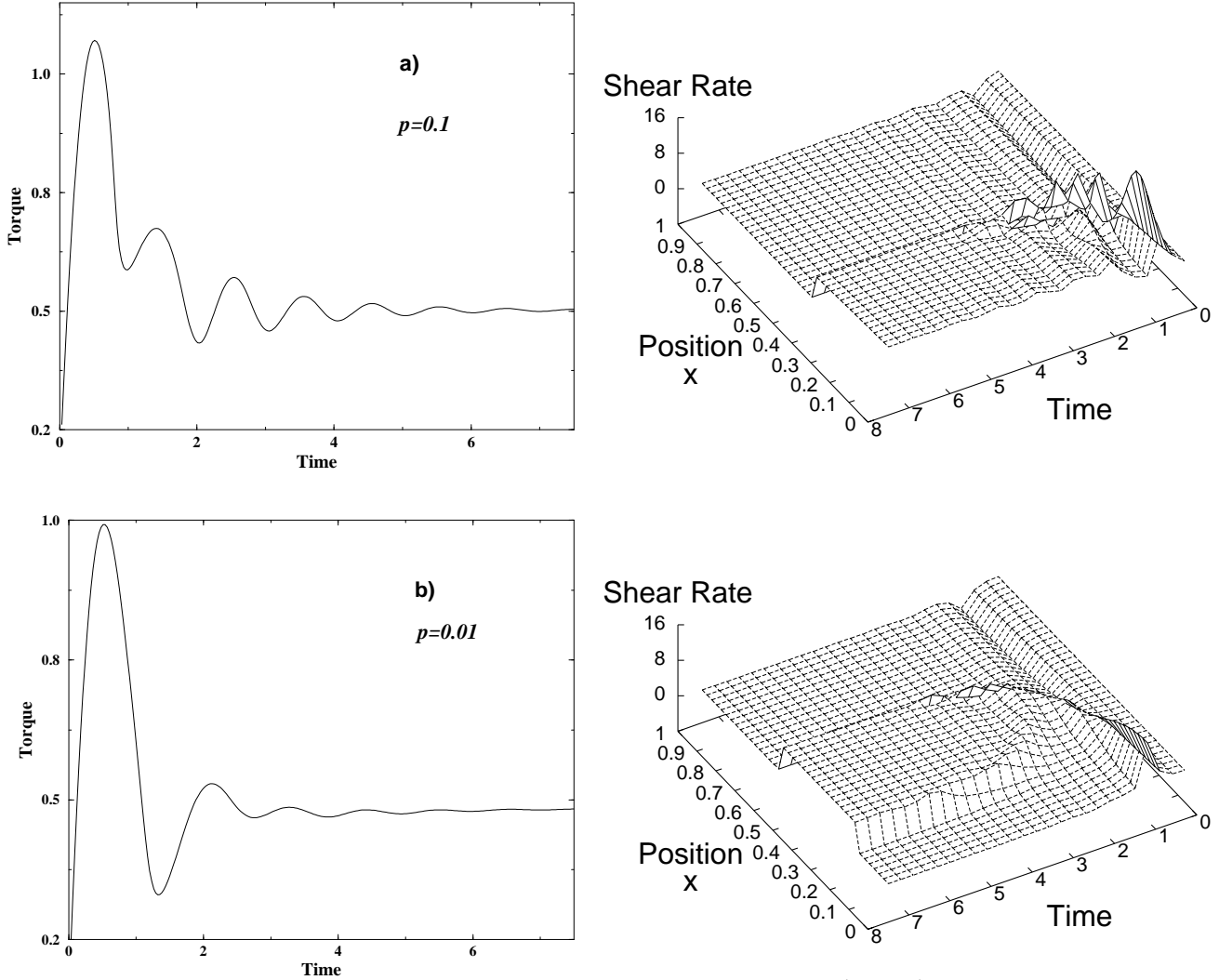}}
\caption{Torque vs. time and strain rate vs. time
  jumping up from rest for ${\cal\hat D }=0, \hat V=3.0,
  \epsilon=0.05$, and a) $p=0.1$, b) $p=0.01$.  Spatial mesh of $200$
  points, time step $0.00015$.}
\label{fig:kinetics1}
\end{figure}

The lengths $t_{i}-t_{i-1}$ of the relaxation intervals were chosen to
be much larger than the characteristic relaxation time for each
velocity change, in order to obtain the steady state at the end of
each interval.  We use an implicit Crank-Nicholson algorithm to
perform the dynamics.  At each time step we must determine the torque
$\hat\Gamma$ through the non-local integral condition,
Eq.~(\ref{eq:Gam}). For comparison, we note that \citeasnoun{greco97}
performed a similar calculation using $\epsilon=0.05$ and $p=0.1,
0.02, 0.01$, and starting from rest.
\end{multicols} \widetext
\begin{figure}[!tbh]
\centering{\includegraphics{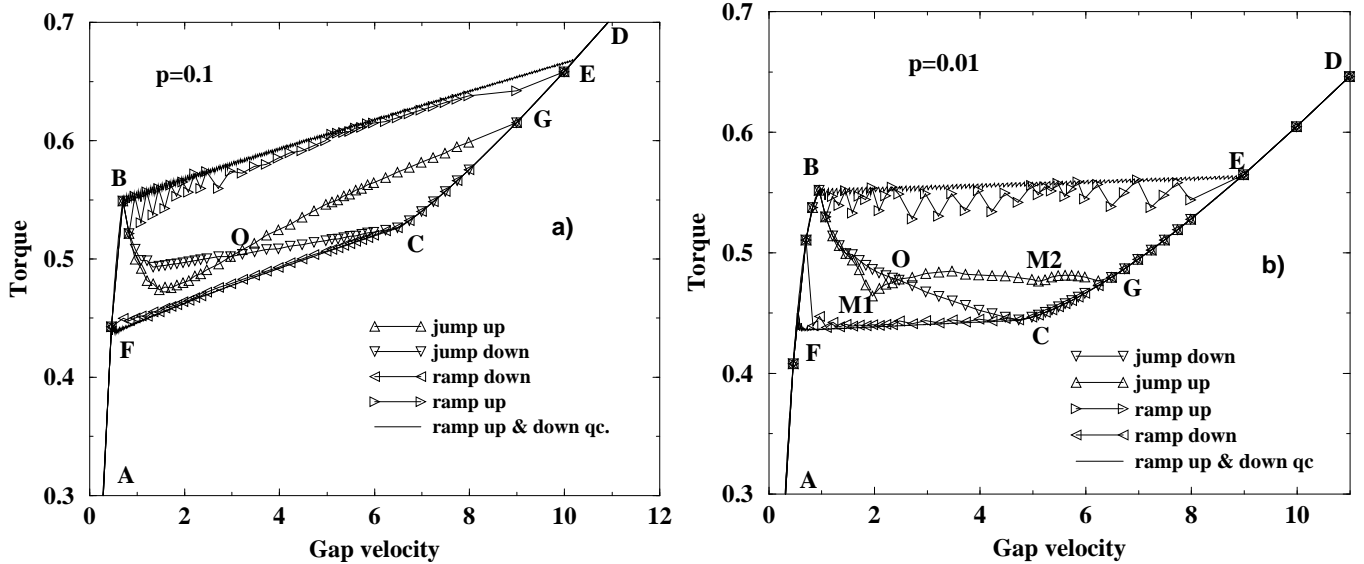}}
\caption{ Torque vs. gap velocity for the four scenarios described in
  the text, for ${\cal\hat D }=0$ and $\epsilon=0.05$.  a) $p=0.1$; b)
  $p=0.01.\, \hat V_{down}=0.0, \hat V_{up}=12.0$, the spatial mesh is
  $200$, the time step is $0.005$, and $10000$ time steps assured
  steady state at each value for $\hat V$. Ramps are shown for both
  large ($d\hat V\gtrsim 0.2$) and quasi-continuous ($d\hat
  V\simeq0.01$, denoted {\bf qc}) velocity changes.}
\label{fig:scenarios}
\end{figure}
\begin{multicols}{2} \narrowtext
  The steady state values of $\hat\Gamma$ at a given gap velocity
  $\hat V$ are history dependent, as shown in
  Fig.~\ref{fig:scenarios}.  For the same gap velocity, steady flows
  obtained via different scenarios differ by the number of bands and
  the proportion of high shear rate band.  For high curvatures steady
  banded flow has two bands for all scenarios, with the high shear
  rate band at the inner cylinder, but the amount of high shear rate
  band for the same gap velocity is maximum (the torque is minimum) on
  ramp-down, minimum on ramp-up, and takes different intermediate
  values for jump-up and jump-down scenarios.  The case of low
  curvature $p=0.01$ provides a surprise in jump-up from rest, also
  noted by Greco and Ball.  Within the interval $1.96 < \hat V < 5.1$
  (between points M1 and M2 in Fig.~\ref{fig:scenarios}b) the steady
  state contains two interfaces, separating an interior high strain
  rate band from two outer low strain rate bands against the cylinder
  walls.  For $0.95 < \hat V < 1.96$ (between points B and M1 in
  Fig.\ref{fig:scenarios}b) we obtain the usual high-low two-band
  sequence, while for $5.1 < \hat V < 6.48 $ (between points M2 and O
  in Fig.\ref{fig:scenarios}b) the two-band sequence is inverted
  (low-high) with the low strain rate band of material at the inner
  cylinder.  The three-band region is separated from the two-band
  region by discontinuities of the slope, $d\hat\Gamma/d\hat V$
  (points M1, M2).  The values of the gap velocity separating these
  different types of steady flow correspond to local minima of the
  torque $\hat\Gamma(\hat V)$ (Fig \ref{fig:scenarios}b).  The time
  development of two-band and three-band profiles can be seen in
  Fig.~\ref{fig:kinetics1}.  The other three scenarios produce the
  normal two band sequence high-low.
 
  Finally, another history-dependent feature occurs on ramp-up and
  ramp-down scenarios. The size of the high shear rate band in steady
  banded flow and therefore the torque value depends on the magnitude
  of the differences $\hat V(t_{i+1}) - \hat V(t_i)$ between succesive
  velocities.  If the velocity is changed quasi-statically
  (infinitesimal changes of $\hat V$), flow curves are smooth and
  correspond to top and bottom jump (the total shear stress at the
  position of the interface corresponds to the maximum and minimum
  values of the local constitutive relation) on ramp-up and ramp-down
  respectively.  Large velocity changes produce spikes in the flow
  curve, both on ramp-up and ramp-down.  The reason for this is the
  pinning of the interface at a stress value which is lower than the
  maximum allowed value or higher than the minimum allowed value, and
  has been also noticed by \citeasnoun{spenley96} for a different
  diffusionless model.
\end{multicols} \widetext
\begin{multicols}{2} \narrowtext
\subsubsection{Noise effects}
\begin{figure}[p ]
\centering{\includegraphics{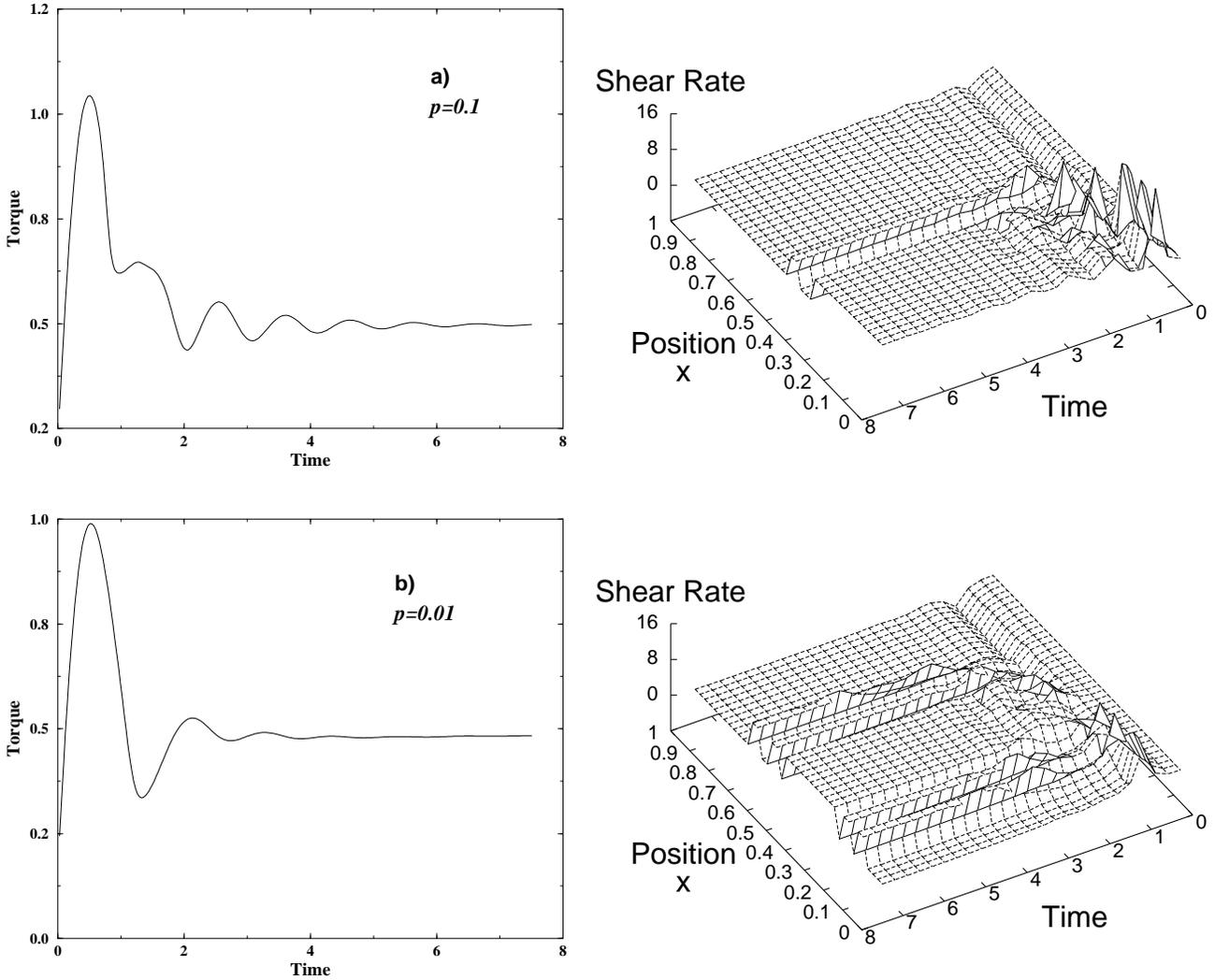}}
  \caption{Torque vs. time and strain rate vs. time
    for ${\cal\hat D }=0, \hat V=3.0, \epsilon=0.05,$ and noisy
    initial conditions, jumping up from rest. a) $p=0.01,
    \eta_{max}=0.01$, time step $0.00015$; b) $p=0.1, \eta_{max} =
    0.05$, time step $0.00015$. Mesh of $200$ space points.}
\label{fig:noise}
\end{figure}

The dynamical selection of steady states in the four scenarios is
unstable with respect to noise in the initial conditions.  To show
this we used perturbed initial
conditions at each new velocity value, in the form
\begin{equation}
  \label{eq:1}
  \Sigma_i^{(j)}(r) = \Sigma_{i-1}^{(j)}(r;\infty) + \eta^{(j)} (r),
\end{equation}
where ${\Sigma_{i-1}^{(j)}}(r;\infty), j=1,3$ are the three
steady-state components of the material stress obtained at step i-1,
and $\eta^{(j)} (r)$ are random, uniformly distributed in the interval
$(-\eta_{max}, \eta_{max})$ and uncorrelated, $\langle\eta^{(j)}(r)
\eta^{(j')}(r')\rangle = 0$ for $r \neq r', j \neq j'$.
Fig.~\ref{fig:noise} shows the result of two-state experiments
($i=1,2$), starting from rest (state $i=1$).  A region in the high
strain rate band close to the position of the interface (that would be
selected without noise) is unstable with respect to noise, and breaks
up into many bands whose widths are limited only by the mesh size.
The size of the unstable region increases with the amplitude
$\eta_{max}$ of the noise and it is smaller for higher curvature
($p=0.1$). Diffusionless dynamics of the banded flow is more stable in
high curvature Couette cell, as suggested by \citeasnoun{greco97}.

\subsubsection{Basins of attraction and continuous degeneracy of
    steady banded flows} 
\begin{figure}[p]
\centering{\includegraphics{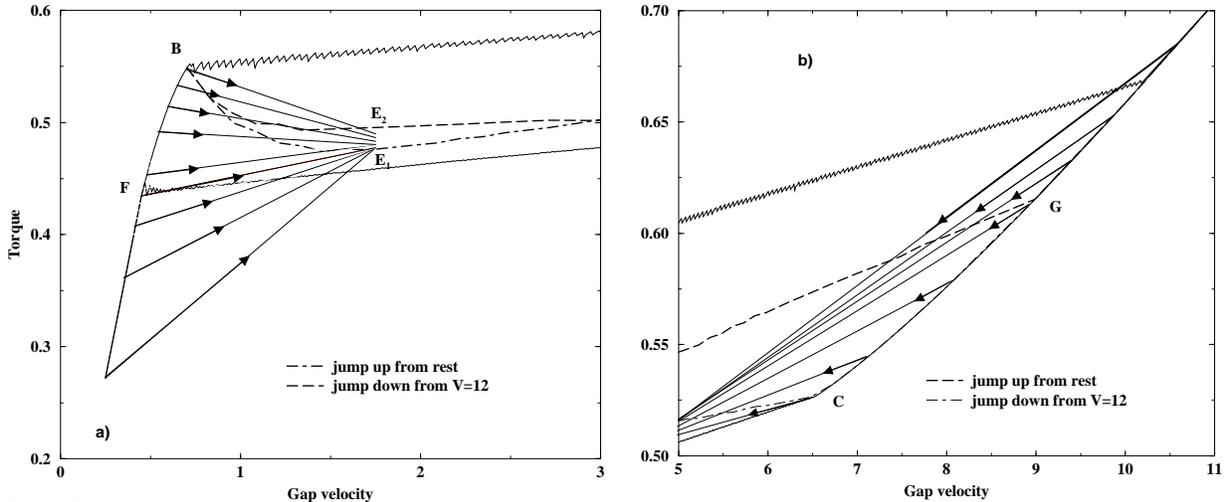}}
  \caption{Structure of basins of attraction. The arrows lines
    connect start and end states (attractors) upon (a) Jumping up from
    the low shear rate band ($\hat V_{end}=1.75$) (b) Jumping down
    from the high shear rate band ($\hat V_{end}=5$).}
\label{fig:attractors}
\end{figure}

  The dynamical equations (\ref{eq:cyl2}) with ${\cal\hat D}=0$
  correspond to an infinite dimensional nonlinear dynamical system.
  The dynamics in the functional space of banded flows has a complex
  structure of attractors (steady banded flows). Without noise,
  certain classes of trajectories, corresponding to initial conditions
  inside basins of attraction, reach the same steady state.  The
  jump-up branch $BG$ (Fig.~\ref{fig:scenarios}) is a set of
  attractors, collecting trajectories starting on the low shear rate
  branch.  For an imposed gap velocity $\hat V_{end}$ in the banded
  regime, jumping up from an initial homogeneous steady state of gap
  velocity $\hat V_{down} < \hat V_F$ (initial state below the point F
  on the flow curve, Fig.~\ref{fig:attractors}a) leads to an identical
  final banded steady state (point $E_1$).  For higher initial gap
  velocities $V_F < \hat V_{down} < V_B$ (initial states between
  points F and B on the flow curve, Fig.~\ref{fig:attractors}a) the
  final torque spans the interval $E_1\!-\!E_2$ between the jump-up
  and jump-down branches (the discreteness in
  Fig.~\ref{fig:attractors} is a finite mesh effect).  The jump-up
  from rest branch is thus the lower limit of a continuous family of
  jump-up attractors.  Similarly, jumping-down from a velocity $\hat
  V_{up} > V_G$ ends on the attractor branch $BC$ which represents the
  upper limit of jump-down attractors which collect trajectories with
  $\hat V_{up} < V_G$ (Fig.~\ref{fig:attractors}).
\end{multicols} \widetext
\begin{multicols}{2}\narrowtext
  
  Despite the interesting structure of these dynamics, experiments on
  wormlike micelles invariably display unique banding flows,
  independent of history and initial conditions.  This is incompatible
  with the numerical results obtained for ${\cal D}=0$; as we shall
  see in the following, the imposition of non-vanishing stress
  diffusion ${\cal D\/}\neq 0$ provides a route for constructing
  unique banding flows \cite{elkareh89,olmstedlu97,lu99}.
\end{multicols} \widetext
\begin{multicols}{2} \narrowtext
\section{Couette flow with diffusion: ${\cal\hat D }\neq
  0$}\label{sec:Dnonzero} 
\subsection{Flow curves and History independence}
\begin{figure}[p ]
\centering{\includegraphics{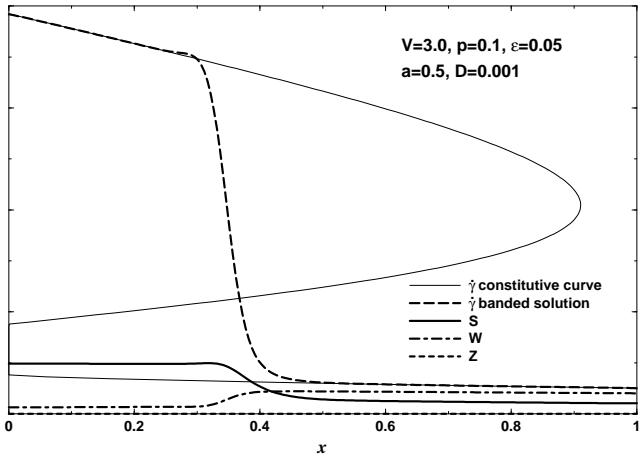}}
\caption{Profiles for a typical steady state banded solution. Shown
  are the local constitutive relation and banded strain rate profile,
  analogous to those in Fig.~\ref{fig:torque}, as well as the
  components $\hat S,\hat W,\hat Z$ of the polymer stress, as a
  function of the spatial coordinate $x$.}
\label{fig:profilesD}
\end{figure}
\begin{figure}[p ]
\centering{\includegraphics{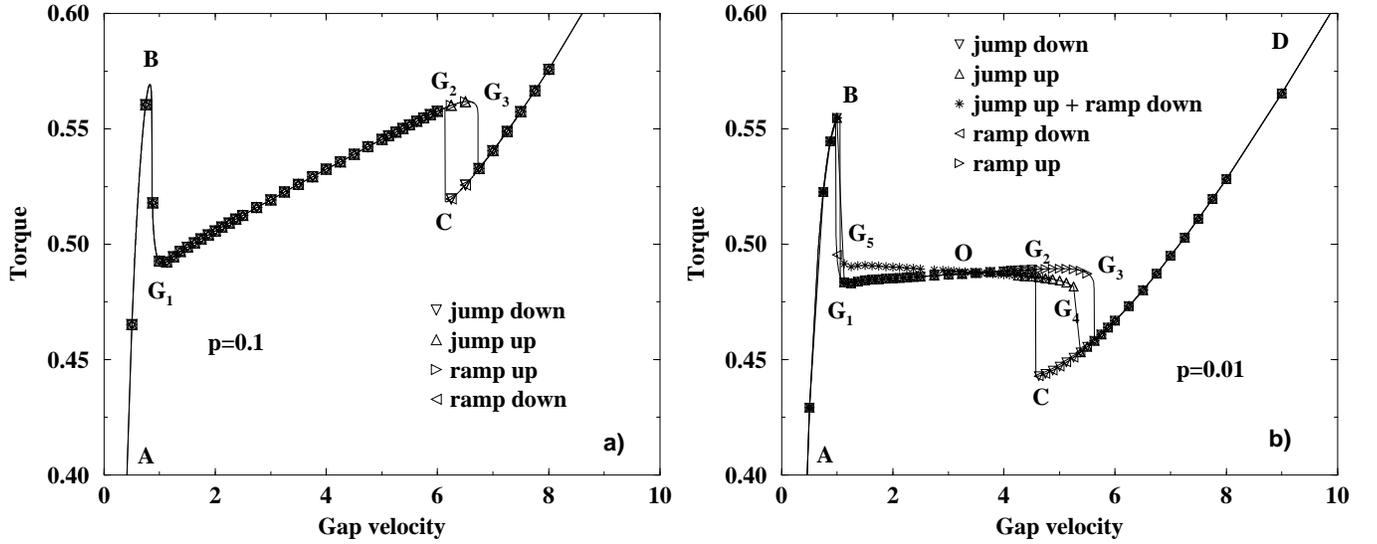}}
\caption{Torque vs. gap velocity for a finite diffusion coefficient
  ${\cal\hat D}=0.005, \hat V_{down}=0.0, \hat V_{up}=12.0$.
  a) $p=0.1, 20000$ time steps, b) $p=0.01, 40000$ time steps. The
  spatial mesh is 400 points, with a time step of $0.005$.}
\label{fig:scenariosD}
\end{figure}

We have repeated the four histories with ${\cal\hat D }\neq 0$. In
this case banded profiles are smooth, as shown in
Fig.~\ref{fig:profilesD}. The local strain rate
$\hat{\dot{\gamma}}(r)$ follows that of either the high or low strain
rate constitutive branches except for the interface region whose width
scales like $\sqrt{\cal\hat D }$.  The flow curves are shown in
Fig.~\ref{fig:scenariosD}; as can be seen, the flow curves are
essentially history-independent. Our main result is that even a small
diffusion term lifts the continuous degeneracy of steady states. We
shall see that this occurs by selecting the value of the total shear
stress at the position of the interface, with a selected stress
$\sigma_{sel}$ independent of imposed gap velocity, interface
position, curvature, or ${\cal D\/}$.

In contrast to planar flow, for which the flow curve has a flat stress
plateau, in cylindrical Couette flow the torque in the banding part of
the flow curve has a slight dependence on the imposed gap velocity
(Fig.~\ref{fig:scenariosD}). This is a consequence of the relation
$\hat \Gamma = \hat{\sigma}_{sel}\, \hat{r}_*^2$, with
$\hat{\sigma}_{sel}$ the selected stress at the interface position
$\hat{r}_{\ast}$, the latter which increases from $1$ to $1+p$ as the
fast-flowing band gradually fills the gap with increasing gap velocity
$\hat{V}$.  The slope of the plateau, $d\hat{\Gamma}/d\hat{V}\sim p$,
vanishes in the planar limit of a thin gap. This provides a second
mechanism for a sloped plateau, in addition to the effects of a
concentration difference between the banding states $\hat\Gamma(\hat
V)$ \cite{schmitt95,olmstedlu97}.

The history independence and uniqueness of banded steady flow is shown
by the coincidence of the segment $G_1\!-\!G_2$ in
Fig.~\ref{fig:scenariosD}a for all scenarios and high curvature
$p=0.1$.  However, some history dependence remains for small
curvatures, as seen by the two plateaus $G_1\!-\!O\!-\!G_3$ and
$G_5\!-\!O\!-\!G_4$ for $p=0.01$ (Fig.~\ref{fig:scenariosD}b).  The
positive slope plateau $G_1\!-\!O\!-\!G_3$ corresponds to the
conventional two-band flow with the high shear rate band at the inner
cylinder, while for the negative slope plateau $G_5\!-\!O\!-\!G_4$ the
band sequence is inverted.  The normal band sequence can be obtained
by jumping up from rest for small gap velocities (between $G_1$ and
$O$) and the inverted sequence can be obtained by jumping up from rest
only for high gap velocities (between $O$ and $G_4$).  In order to
scan the entire lengths $G_1\!-\!G_3$ and $G_4\!-\!G_5$ of the
plateaus, ramping is necessary after jumping up, as shown in Fig.
\ref{fig:scenariosD}.  We have checked that both band sequences are
linearly stable by considering small perturbations of the interface
position and shear rate.  Neverthess, as discussed in
\cite{radulescu99a}, the inverted band sequence is only metastable and
nucleation processes may change the band sequence.  An inverted band
is possible when both the high and low strain rate branches of the
constitutive curve span the entire gap, as in Fig.~\ref{fig:torque}c.
There has been no experimental evidence yet for the existence of the
metastable inverted band sequence, but this could be because either
most experiments are performed by jumping up at low $\hat{V}$, or
apparatus noise and interface oscillations destabilize the inverted
band sequence.

\end{multicols} \widetext
\subsection{Selection mechanism}
\begin{figure}[p ]
\centering{\includegraphics{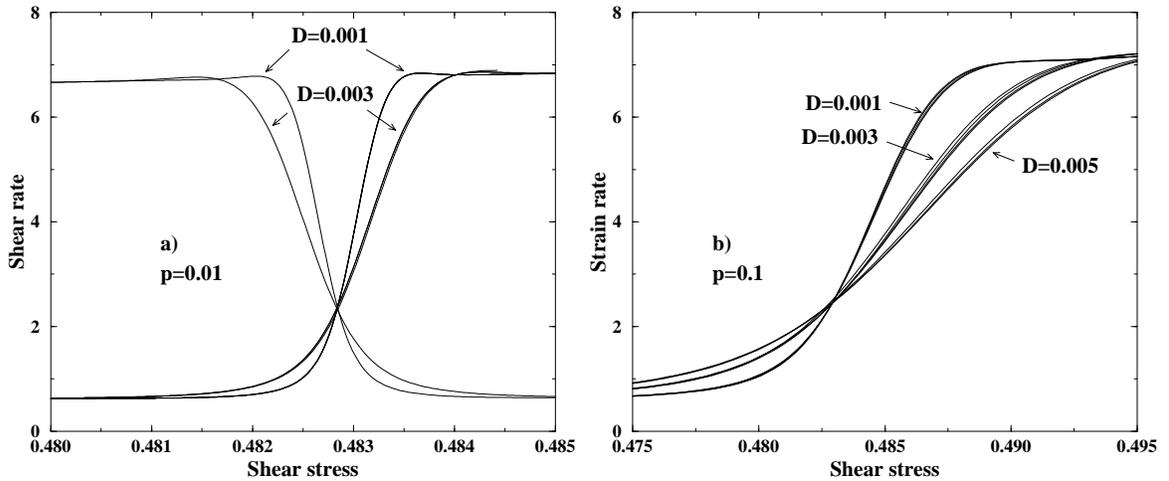}}
  \caption{Several interface profiles in the representation 
    $\hat{\dot{\gamma}}(\hat\sigma)$. (a) two different sequences of
    the bands (high-low and low-high), $p=0.01, \hat V = 1.5, 2.5,
    3.5, 4.75$ and ${\cal \hat D} = 0.001, 0.003$; (b) $p=0.1$ and
    ${\cal \hat D} = 0.001, 0.003, 0.005$ and $\hat V = 2.0, 3.0, 4.0,
    5.0, 6.0$. For a given ${\cal\hat{D}}$ the profiles for different
    $\hat{V}$ differ imperceptibly.}
\label{fig:selection}
\end{figure}
\begin{multicols}{2} \narrowtext
  
  The key feature of stress selection by diffusion is that a
  stationary interface is stable only at the unique position
  $r_{\ast}$ inside the gap corresponding to a selected value
  $\hat{\sigma}_{\ast}$ (as long as $r_{\ast}$ is a distance from the
  walls which is large compared to the interface width).  A rigorous
  proof of this property does not exist in the case of J-S model, but
  justifications can be found in \citeasnoun{lu99}, and
  \citeasnoun{radulescu99a}.  To check this we have represented the
  steady shear rate profiles as a function of $\hat\sigma = \hat\Gamma
  / \hat{r}^2$ instead of $\hat{r}$ (Fig.~\ref{fig:selection}).  In
  this representation different banded profiles for different values
  of $\hat V$, and hence $\hat{r}_{\ast}$, practically coincide for
  the same value of ${\cal\hat D }$. Profiles obtained for different
  values of $\cal\hat D$ or with different band sequences intersect at
  the same stress $\hat\sigma^{\ast}\equiv\hat\sigma^{\ast}_{sel}$
  within $0.01\%$, which we define as the selected stress (see
  Fig.~\ref{fig:selection}).  This selected stress is independent of
  the cell curvature, the diffusion coefficient $\cal\hat D$, and the
  gap velocity $\hat{V}$.  For example, for $\epsilon=0.05, a=0.5$ we
  find a selected stress $\hat\sigma^{\ast}_{sel}\simeq 0.4828$, which
  is almost identical to the selected stress for the planar JS model
  for the same viscosity ratio, $\hat\sigma^{\ast}_{planar}=0.4827$
  \cite{lu99}.

  Hence, for interface widths much smaller than the radii of curvature
  or cylinder gap, the flow is essentially homogeneous on the scale of
  the interface, and the interface migrates to that position in the
  cylinder for which the selected stress $\sigma^{\ast}_{sel}$
  obtains.  The selection of $\hat\sigma^{\ast}_{sel}$ provides a
  relation $\hat\Gamma / \hat{r}_*^2 =\hat\sigma^{\ast}_{sel}=const.$
  between the torque and the interface position, thus fixing
  $\hat\Gamma$ for a given gap velocity $\hat V$ and band sequence.
  The magnitude of $\hat{\cal D\/}$ has negligible effect in the thin
  interface limit.
  
\begin{figure}[!tbh]
\centering{\includegraphics{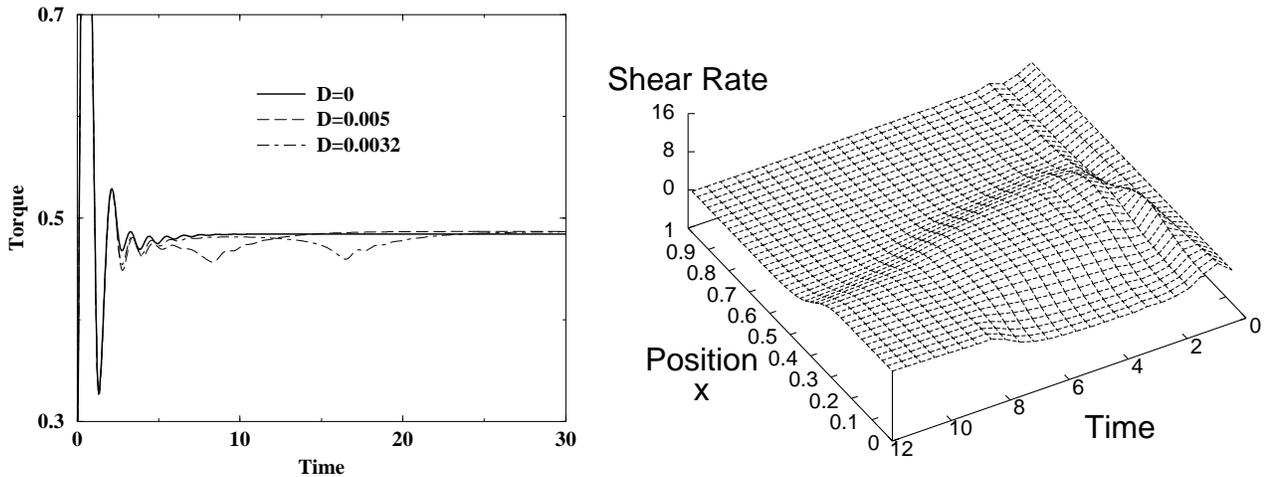}}
\caption{(left)  Strain rate vs. time 
  for $p=0.01, \hat{\cal D}=0.0, 0.0032,$ and $0.005$ and $\hat
  V=3.0$.  (right) Torque vs. position for ${\cal D\/}=0.005$.
  Elimination of the interface occurs at a local minimum (in time) of
  the torque. Mesh of $200$ space points, time step $0.00024$, for
  startup from rest.}
\label{fig:kinetics2}
\end{figure}

  Stress selection has interesting consequences for the kinetics, one
  of which is illustrated in Fig. \ref{fig:kinetics2}.  For flows
  jumping up from rest with gap velocities in the banding regime, the
  kinetics until an interface is well-developed are the same as the
  kinetics for ${\cal\hat D } = 0$, as seen by coincidence of the
  solid and broken curves in Fig.~\ref{fig:kinetics2}.  For ${\cal
    \hat D } = 0$ the kinetics ends with the {\em formation\/} of one
  or several interfaces, while in the presence of diffusion a second
  stage follows in which the interface migrates to its equilibrium
  position at the selected stress. For cylindrical Couette flow the
  stress is a monotonic function of position, so this equilibrium
  position is unique and steady banded flow has only one interface. If
  multiple interfaces form the ``excess'' interfaces will eventually
  be expelled at the walls. This is the case for low curvature
  (Fig.~\ref{fig:kinetics2}).  The three band solution obtained when
  ${\cal \hat D } = 0$ (Fig.~\ref{fig:kinetics1}) becomes a transient
  for ${\cal \hat D } \neq 0$.  For low gap velocities the interface
  closest to the inner cylinder is expelled and the normal sequence of
  bands is obtained at stationarity. Note that the expulsion time
  increases for decreasing ${\cal\hat D\/}$ (compare ${\cal\hat
    D\/}=0.0032$ and ${\cal\hat D\/}=0.005$).  The inverted sequence
  results from an initial three-band transient at higher velocities
  when the interface closest to the outer cylinder is eliminated. The
  timescale of the interface displacement is long for small ${\cal\hat
    D}$, so a three-band transient could have a very long life and be
  mistaken for steady flow.
\end{multicols}  \widetext
\begin{multicols}{2}\narrowtext
\subsection{Hysteresis}
\begin{figure}[p]
\centering{\includegraphics{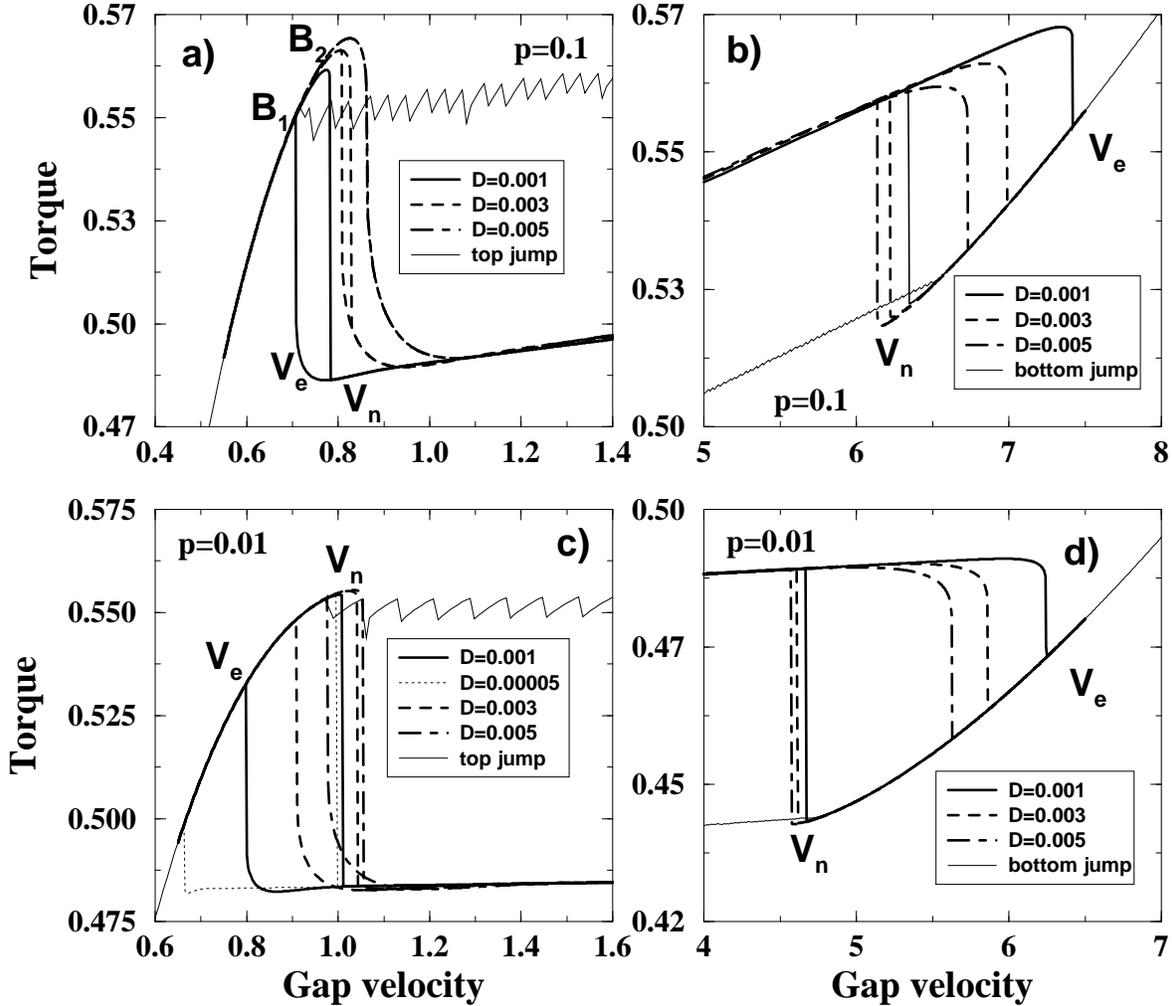}}
\caption{Small velocity step ramping up and down ($d\hat V=0.002$) 
  in the hysteresis regions, with diffusion.  a) $p=0.1$ left end of
  plateau; b) $p=0.1$ right end of plateau; c) $p=0.01$, left end of
  plateau, d) $p=0.01$, right end of plateau.  The spatial mesh is 200
  points, with a time step of $0.005$.  In all cases the diffusionless
  top and buttom jump are shown, respectively. }
\label{fig:hysteresis}
\end{figure}

Ramping up and down at the extremities of the plateaus define
different paths, as shown in Fig.~\ref{fig:hysteresis}: the low strain
rate branch (Fig.~\ref{fig:hysteresis}a,c) or high strain rate branch
(Fig.~\ref{fig:hysteresis}b,d) and the banded plateaus both possess
locally stable steady state flows for a narrow range of strain rates.
Upon ramping up from zero the high shear rate band begins to form at a
gap velocity $V_n$ (point $B_2$ in Fig.~\ref{fig:hysteresis}a) which
is slightly different than the top-jump value.  $V_n$ converges to the
top-jump value when ${\cal D} \rightarrow 0$.  On ramping back down
the interface between bands is eliminated when it reaches a distance
from the wall of order its width (point $B_1$). The difference between
the corresponding gap velocity $V_e$ and the extrapolation to the low
strain rate branch scales like $\sqrt{\cal D}$.  Furthermore the width
of the hysteresis loops, $|\hat V_{n} - \hat V_{e}|$, increases with
decreasing $\cal\hat D$, consistent with the banded plateau remaining
until the interface ``touches'' a cylinder wall. For large enough
${\cal D\/}$ hysteresis vanishes.

For small ${\cal D\/}$ the hysteresis loop at the low shear rate end
of the plateau contains the metastable part $B_1-B_2$ of the low shear
rate branch. This is reminiscent of the experimental results of
\citeasnoun{grand97}: under controlled strain rate conditions they
found a stress (analogous to $B_1$ in Fig.~\ref{fig:hysteresis}a)
below which the system remained homogeneous, and above which the high
strain rate band nucleated. Similar phenomena are expected at the high
shear rate end of the plateau, but an experimental study of this
region raises difficult problems because of various noise sources.
\end{multicols} \widetext
\begin{multicols}{2}\narrowtext
\section{Summary}\label{sec:sum}
We have studied shear banding in the local (${\cal D\/}=0$)
Johnson-Segalman model in cylindrical Couette flow under controlled
strain rate conditions.  The inhomogeneous flow conditions of the
Couette geometry generate history dependent steady states with a
single interface of zero width. We have outlined some of the behavior
under various flow histories, and identified families of attractors
which govern the behavior in the absence of non-local terms in the
dynamics.

To recover a unique stress and the history independence seen in
experiments, it is necessary to augment local constitutive equations
with non-local information.  We have added a simple diffusion term and
shown that the banding profile is, in the main, history independent.
The steady state flow curves we have found are consistent with the
growing body of work on wormlike micelles; moreover, regions of the
flow curves suggest the analogs of metastability, supercooling, and
nucleation and growth, and are suggestive of recent experiments on
wormlike micelles \cite{berret94b,Berr97,grand97}.  The flow curves
have a selected banding torque which increases with increasing gap
velocity, with a more pronounced increase for more highly curved
Couette geometries.

The diffusion term represents a singular perturbation which allows the
interface to only lie at that position for which the shear stress has
a well defined selected value. In the Couette geometry this position
in unique, and multi-interface steady state solutions are prohibited.
n For flatter geometries a metastable inverted band sequence exists,
with the high strain rate phase near the outer cylinder.  However,
there is as yet no accepted theory for diffusion terms in wormlike
micelle solutions, and nor have they been measured. Our results
suggest further interesting effects to study, such as band nucleation
and expulsion, interface motion, and kinetics, which we hope to
address in the future. For example, $\cal\hat D$ controls interface
speed, which thus provides a possible experimental measure of this
unknown quantity.

\emph{Note Added---}After this manuscript was completed we learned of
recent work by \citeasnoun{yuan99}, in which the JS model was
supplemented with a diffusion term in the strain rate (effectively a
non-local viscosity), and calculations were performed using a 2+1D
Eulerian-Lagrangian technique in planar shear flow. Yuan also found
unique stress selection in the presence of a diffusion term, but the
results with no diffusive term still gave unique (and different)
stress selection.  This indicates that either the numerical
discretization technique introduces spurious non-local terms into the
dynamics, or that the larger fluctuation phase space of higher
dimensions somehow stabilizes a unique stress in the absence of
diffusion terms. The latter possibility appears to contradict
\citeasnoun{renardy}.

\noindent{\bf Acknowledgments}
We thank J-F Berret, G Porte, J-P Decruppe, Tom McLeish and Scott
Milner for helpful discussions, and acknowledge funding from St.
Catharine's College, Cambridge and the (Taiwan) National Science
Council (NSC 87-2112-M-008-036) (CYDL); and EPSRC (GR/L70455) (OR,
PDO).

\end{multicols}
\end{document}